**Evidence of length scale effect in contact electrification in conducting thin film heterostructures**

Paul C. Lou[1], Ravindra G. Bhardwaj[2], Anand Katailiha[1], W.P. Beyermann[3] and Sandeep Kumar[4*]

[1] Department of Mechanical Engineering, University of California, Riverside, CA, USA

[2] Department of Mechanical Engineering, Birla Institute of Technology and Science Pilani, Dubai Campus, Dubai, the United Arab Emirates

[3] Department of Physics and Astronomy, University of California, Riverside, CA, USA

[4] Independent researcher, Kurukshetra, India

[*] Corresponding author

Email- sandeep24jan@gmail.com

Abstract

Contact electrification between two conducting materials is expected to exhibit length scale effect because the screening effect will diminish in conductors as a function of material dimensions. As a consequence, the interfacial charge accumulation will diffuse away from interface/surface to a critical penetration depth as a function of material dimension. This work experimentally demonstrates the length scale effect in a permalloy and degenerately doped p-Si heterostructure system due to the flexoelectricity mediated contact electrification. The contact electrification induced interlayer charge transfer is observed through the whole thickness in case of 400 nm thick p-Si samples. Whereas, the charge carrier diffuses to a depth of 51 nm from the interface in case of 2 μm thick Si. The length scale effect also leads to metal-insulator transition in p-Si layers in both cases. These results present a new opportunity to tailor the physical properties in conducting materials using contact electrification.

When two similar or dissimilar materials are brought into contact with each other and then separated, a charge with opposite sign will appear on the contacting surfaces[1-6]. This behavior is called contact electrification (CE) and is known since ancient times. The contacting materials can be any type such as metal, insulator, semiconductor or polymer. In case of dissimilar metals, the charge accumulation at the surfaces occurs until their Fermi levels are coincident[3, 5, 7]. Recent studies show that the charge accumulation will not remain at the surface but it will diffuse from surface to interior of a conducting material at nanoscale[8, 9], which give rise to electronic dynamical multiferroicity[10, 11], spin dependent response[12-15] and magnetic behavior in non-magnetic Si[16] (degenerately doped). Based on these reports, we postulate that the CE behavior in conducting thin films may exhibit length scale effect.

The CE will give rise to a surface/interfacial electric field. In bulk conducting materials, the absolute number of free charge carrier is large and the surface/interfacial electric field will be screened. The charge will remain at the surface/interface without affecting the interior as shown in Figure 1 (a). As the dimensions of the contacting materials are reduced, a critical length scale will be reached where surface/interfacial electric field due to CE will not be completely screened. As a consequence, the surface/interfacial charge will diffuse to the interior of the conducting material as shown in Figure 1 (a). Further reduction in the dimension will lead to negligible electrostatic screening from the charge carriers, which will lead to charge transfer through the thickness of the thin film as shown in Figure 1 (a). This behavior is similar to length scale effect observed in other physical (electrical, mechanical, optical and thermal) properties[17]. In this work, we experimentally demonstrate the length scale effect in flexoelectricity mediated CE[8, 18, 19] in permalloy (Py)/p-Si (degenerately doped) thin film heterostructures. The interlayer charge transfer from CE diffuses primarily to a depth of ~51 nm in case of 2 μm thick p-Si layer whereas interlayer charge transfer is across the full thickness of 400 nm of p-Si layer.

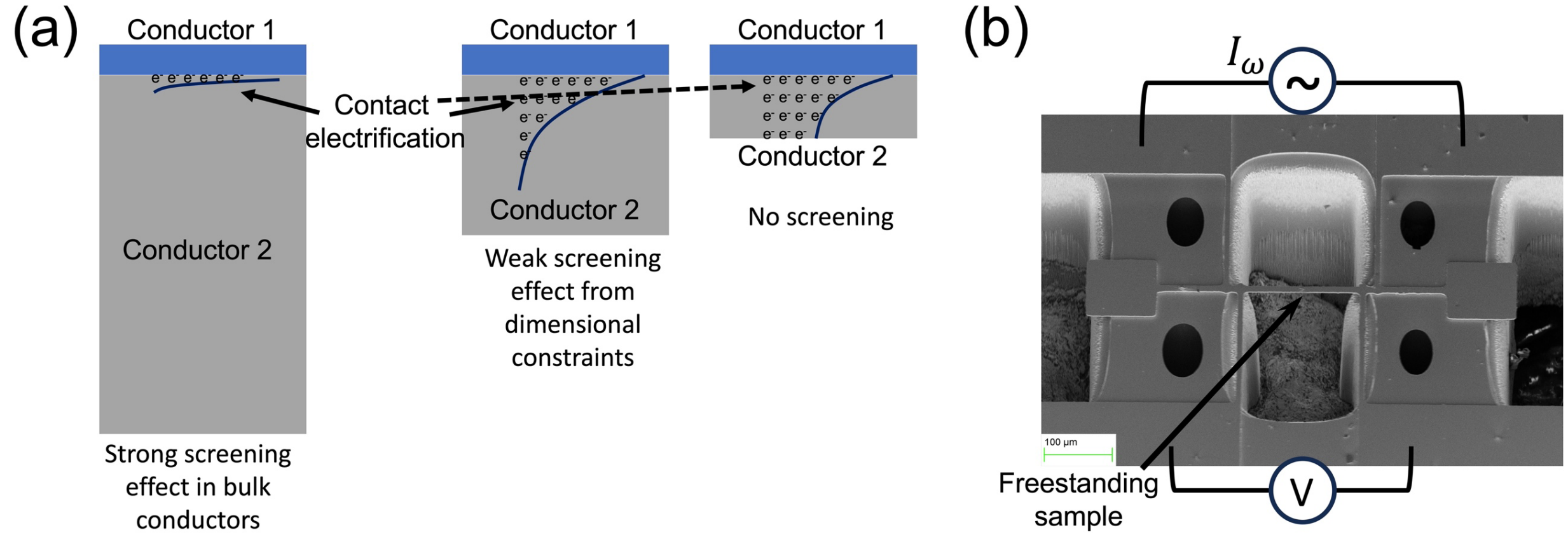

Figure 1. (a) A schematic showing the CE behavior expected due to length scale effect and resulting change in the screening effect, and (b) a representative scanning electron micrograph showing the freestanding sample structure and measurement scheme.

While the traditional CE behavior is studied using contact and separation between two materials but charge accumulation takes place during the contact[3, 18] itself. Hence, separation of the contacting surfaces is not an essential requirement for measuring the CE response. Therefore, we studied the charge transfer behavior experimentally in a heterostructure sample without separation of the thin films unlike conventional CE experimental setups. In the experimental setup, the heterostructure samples are made freestanding as shown in Figure 1 (b) and, as a consequence, they will deform (buckle) due to residual stresses from thin film processing. Hence, the CE in these heterostructures is achieved using strain gradient mediated flexoelectricity[8, 18, 19].

To demonstrate the length scale effect in the CE in conducting materials, we fabricated two sets of freestanding heterostructure samples having following two configurations: 1. Pd (1 nm)/ Py (25 nm)/MgO (1.8 nm)/$SiO_2$ (native 3.8 nm)/p-Si (400 nm) and 2. Pd (1 nm)/ Py (25 nm)/MgO (1.8 nm)/$SiO_2$ (native 3.8 nm)/p-Si (2 μm). The details of the sample fabrication and device structure are available in ref. [14, 15]. The sample length for both the configuration is 160 μm. The oxide layers inhibited the diffusion and alloying. The thickness of the Py layer is kept constant and p-Si layer thickness is varied to uncover the length scale effect in the p-Si layer. The electrical resistivity value of the Py is two orders of magnitude smaller than the degenerately doped p-Si. This difference in electronic properties makes them suitable for CE studies in conducting materials

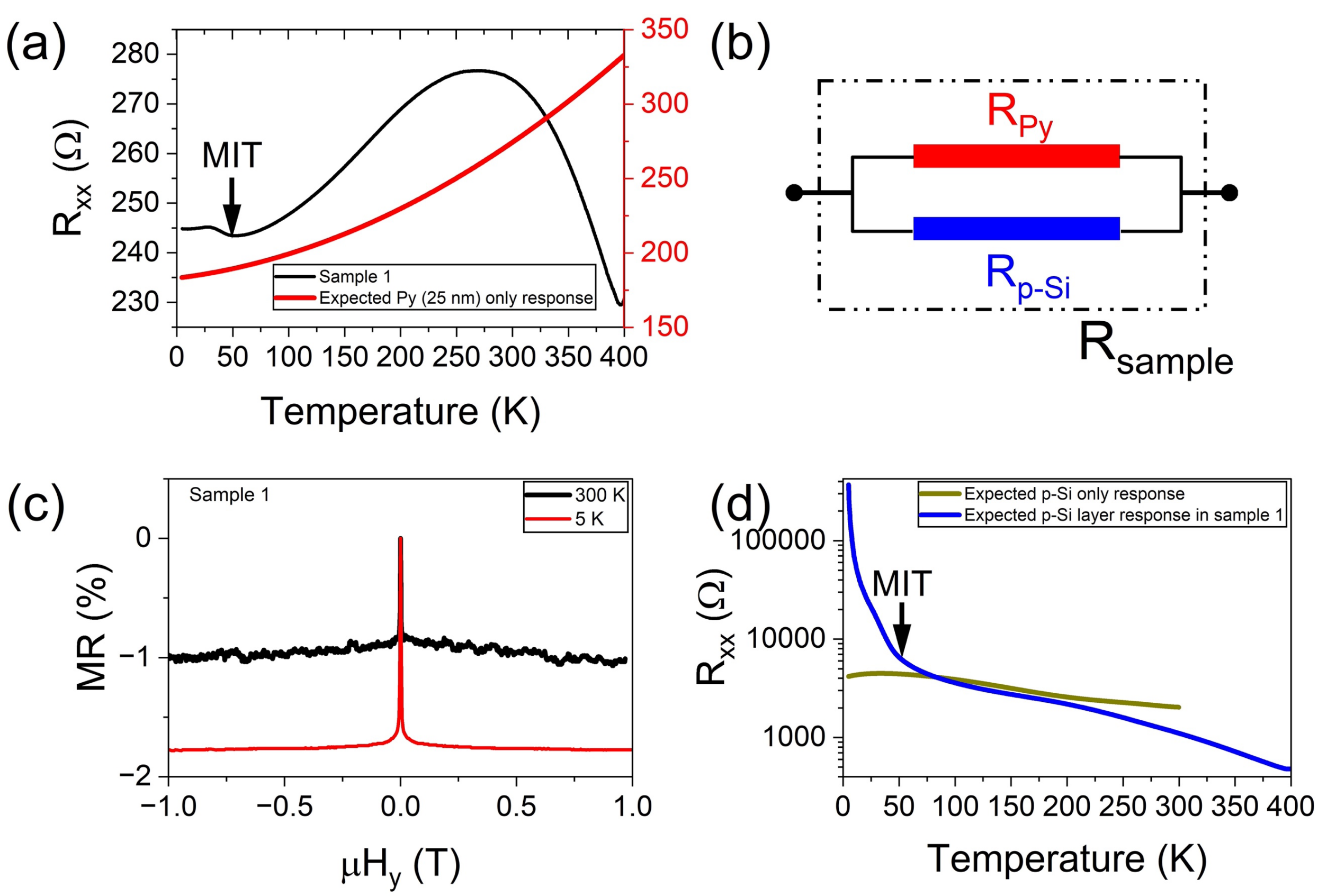

Figure 2. (a) The resistance response as a function of temperature in Py (25 nm)/MgO (1.8 nm)/p-Si (400 nm) sample (sample 1) and control Py response (estimated) from 400 K to 5 K. (b) Schematic showing the Py and p-Si layers being resistors in parallel that give rise to heterostructure sample resistance. (c) the magnetoresistance in sample 1 as a function of transverse in-plane magnetic field from 1 T to -1 T at 300 k and 5 K. And (d) the resistance response of a control p-Si thin film from 300 K to 20 K and p-Si layer in the sample 1 from 400 K to 5 K.

First, we took a sample (sample 1) from first configuration and measured the resistance response as a function of temperature. The resistance response was measured using lock-in technique in a Quantum Design Physical Property Measurement System (PPMS) from 400 K to 5 K at 2 mA and 7 Hz of current bias as shown in Figure 2 (a). The resistance decreases as the temperature is reduced from 400 K but it starts to increase below 396.5 K. The resistance response shows a peak at 271.6 K where the sample resistance is 276.76 Ω as shown in Figure 2 (a). The resistance response shows an increase in resistance followed by a plateau below 52 K as shown in Figure 2 (a). In order to understand the resistance behavior, we needed to extract the responses from Py and p-Si layers. In the heterostructure sample, the conducting Py and p-Si layers can be considered as two resistors ($R_{Py}$ and $R_{p\text{-}Si}$) in parallel configuration giving rise to the total sample resistance ($R_{sample}$) as shown in Figure 2 (b). The temperature (T) dependent resistivity of the Py can be described by the following equation[9, 14]:

$$\rho_{Py}(T) = 3.289 \times 10^{-7} + 1.65 \times 10^{-10} T + 1.269 \times 10^{-12} T^2 \quad (1)$$

The resistance profile of the control Py (25 nm) thin film is calculated using equation 1 and as shown in Figure 2 (a). The measured resistance in the heterostructure sample at 5 K is 244.83 Ω, which is larger than the expected resistance of Py (182.6 Ω) at 5 K[9, 14]. In parallel resistor configuration, the sample resistance cannot be larger than any individual resistances, which means that the sample resistance cannot be larger than the resistance of Py or p-Si layer individually. The measured resistance can only occur if either the p-Si layer or the Py layer is completely insulating at 5 K. In order to verify this postulate, we measured the magnetoresistance (MR) in sample 1 for a transverse in-plane (y-axis) applied magnetic field from 1 T to -1 T at 300 K and 5 K as shown in Figure 2 (c). The MR response increases when the temperature is reduced to 5 K from 300 K as shown in Figure 2 (c). The presence of MR response at 5 K suggests that Py layer is still conducting and the residual resistance is due to Py only. As a consequence, the p-Si layer must be the insulating layer.

Next, we needed to extract the response of p-Si layer in the sample 1 and then compare it to the control p-Si thin film response. To do that, we needed the Py resistance response in the heterostructure sample 1. Based on the residual resistance at 5 K, we calculated a multiplication factor of ~1.335 for control Py thin film response that will give rise to a residual resistance of ~ 244.83 Ω at 5 K. Using this multiplication factor, the Py resistance response in the sample 1 is estimated. It is then subtracted from the sample 1 resistance response (parallel resistor configuration) to extract the resistance response of the p-Si layer in the heterostructure sample, which is shown in Figure 2 (d). The temperature dependent resistivity for a control p-Si (400 nm) thin film sample is extracted (using polynomial fit) from a previously reported data on a sample from the same wafer set as sample 1. Using the resistivity data, we estimated the resistance response for a control p-Si thin film in the absence of any effect from Py layer as shown in Figure 2 (d).

The comparison of resistance behavior as a function of temperature between control p-Si thin film and p-Si in the sample 1 shows that the resistance of the p-Si layer in the sample 1 is an order of magnitude smaller at higher temperatures. Further, the slope of both the responses is similar between 200 K and 75 K even though the resistance of p-Si in sample 1 is lower. At temperature below 75 K, the control p-Si response plateaus. However, the resistance behavior of p-Si layer in sample 1 exhibits a sharp change in slope below ~52 K as shown in Figure 2 (d). This change in slope make the p-Si insulating and is referred as a metal-insulator transition (MIT) as shown in Figure 2 (a) and (d).

At 300 K, the estimated resistance of control p-Si is 2032.3 $\Omega$ whereas the resistance of p-Si in sample 1 is estimated to be 1107.4 $\Omega$. In a Si wafer, small variations in the resistivity can arise due to inhomogeneity of doping. However, the 45% difference in resistance between two Si devices cannot arise due to doping variation. Such a large change in the resistance can only arise from change in charge carrier concentration since mobility is not expected to change significantly. Hence, the charge carrier concentration in the p-Si layer in the sample 1 must be 45% higher at $1.26\times10^{19}$ $cm^{-3}$ at 300 K as compared to ~$8.73\times10^{18}$ $cm^{-3}$ at 300 K[14] in the control p-Si thin film. This increase in charge carrier concentration cannot arise spontaneously without any external stimulus. In the present study, the Py resistance in the sample 1 is also 33.5% larger than the control Py thin film response. An increase in the Py resistance and decrease in the p-Si resistance in the sample 1 can be reconciled using charge carrier transfer from Py layer to p-Si layer. This interlayer charge transfer arises due to strain gradient and resulting flexoelectric effect mediated CE. The excess charge carrier in p-Si layer induces strong electron-electron repulsion. In addition, the charge carrier transfer from Py will also induce ferromagnetic proximity effect in the p-Si layer. This gives rise to Mott MIT observed at ~52 K[9]. In this experiment, the charge carrier transfer is across the thickness (400 nm) of the p-Si layer. This behavior arises due to absence of screening effect at this length scale as hypothesized in Figure 1 (a).

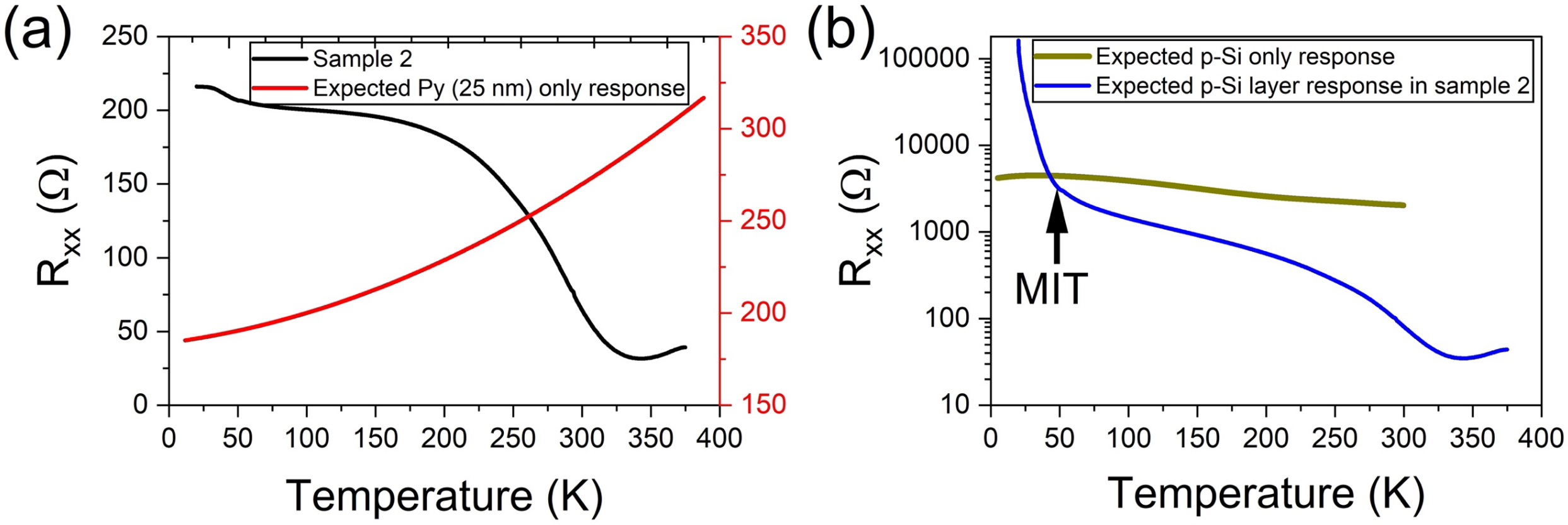

Figure 3. (a) The resistance response as a function of temperature in Py (25 nm)/MgO (1.8 nm)/p-Si (400 nm) sample 2 and control Py thin film response (expected) from 375 K to 20 K. And (b) the resistance response of a control p-Si thin film from 300 K to 20 K and p-Si layer in the sample 2 from 375 K to 20 K.

To reinforce the observed behavior, we repeated the experiment with a second sample (sample 2) with same configuration as sample 1. In the sample 2, we measured the resistance between 375 K to 20 K at 290 μA and 7 Hz of current bias as shown in Figure 3 (a). The resistance initially decreases but, then, starts to increase after 343 K. Similar to sample 1, the sample 2 also exhibits sudden change in resistance at ~52 K and plateau after that. Based on the control Py response, the resistance of the Py layer should be ~193.6 Ω at 20 K as shown in Figure 3 (a). However, the actual measured resistance in the sample 2 is ~ 216.5 Ω at 20 K. Using the methodology similar to the sample 1, we estimated a multiplication factor of 1.112 for the Py response in sample 2. Using the multiplication factor, the response of the Py layer is subtracted from the measured response in sample 2 to extract the resistance response of p-Si layer in the sample 2 as shown in Figure 3 (b). The expected resistance response from the control p-Si sample is also shown in Figure 3 (b). The p-Si layer resistance in the sample 2 is found to be two orders of magnitude smaller than the control p-Si resistance at higher temperatures. Similar to sample 1, the Py layer resistance increased and p-Si layer resistance decreased in the heterostructure sample 2 as compared to control Py and p-Si thin films. This behavior is attributed to the interlayer charge transfer due to flexoelectricity mediated CE similar to the sample 1. The p-Si layer response in sample 2 also exhibits a change in slope at ~52 K, which is attributed to MIT similar to sample 1 even though the resistance response as a function of temperature is significantly different in sample 1. Both samples show MIT across the 400 nm of p-Si. Hence, the flexoelectricity mediated CE will lead to charge transfer across the thickness in p-Si due to lack of screening effect if the thickness of the degenerately doped p-Si is of the order of 400 nm.

Next, we took a sample (sample 3) from second configuration (Pd (1 nm)/ Py (25 nm)/MgO (1.8 nm)/$SiO_2$ (native 3.8 nm)/p-Si (2 μm)) having 2 μm p-Si layer. We kept the Py layer thickness same as sample 1 and 2 but increased the thickness of p-Si layer since we expected the screening effect to be stronger at 2 μm thickness. As a consequence, the interlayer charge transfer from flexoelectricity mediated CE will penetrate a depth that smaller than the p-Si thickness. We measured the longitudinal resistance using lock-in technique for an applied current bias of 100 μA and a frequency of 7 Hz. The resistance response as a function of temperature from 400 K to 5 K (cooling rate of 0.4 K/min) is shown in Figure 4 (a). Then, we did a similar measurement at current bias of 900 μA as shown in Figure 4 (a). Usually, the larger current bias will heat the sample and increases the resistance of the sample. However, the resistance response in sample 3 is lower at 900 μA as compared to response at 100 μA of current bias as shown in Figure 4 (a). While the larger current will heat the sample but it also causes Joule heating[9, 15] mediated thermal mismatch stresses due to different thermal properties in each layer of heterostructure. As a consequence, the strain gradient in a freestanding sample changes due to change in electrical current across the sample. The thermal mismatch stresses and strain gradient increase the flexoelectricity mediated CE response in the heterostructures[8]. The CE leads to charge carrier transfer from Py to p-Si; as demonstrated in sample 1 and 2 as well. The charge carrier mobility is higher in p-Si and, as a result, the resistance of the sample decreases. The observed behavior has been demonstrated previously[10, 11, 14]. This behavior is also similar to electrostatic/ionic gating[20-22] of materials.

We observed an increase in the resistance below 260 K in case of 900 μA current bias measurement as shown in Figure 2 (a). A polynomial fit into the resistance data above and below this transition clearly demonstrate the increase in resistance as shown in Figure 4 (a) inset. To further ascertain the behavior, we took the derivative of the both resistance responses. The derivative of the resistance response measured for 900 μA current bias clearly shows a valley at ~252.5 K as shown in Figure 4 (b). This can be considered a phase transition and, specifically, a metal to insulator transition (MIT). We attribute the MIT at 252.5 K to charging due to CE (enhanced electron-electron repulsion) and ferromagnetic proximity effect due to Py layer, which is expected to be a Mott MIT similar to sample 1 and sample 2. However, only a small layer of p-Si exhibits MIT unlike the sample 1 and sample 2, which is related with the penetration depth of CE.

To estimate the penetration depth and thickness of the MIT layer, we analyzed the resistance data. The resistances of sample at 300 K are 164.38 Ω and 161.37 Ω for 100 μA and 900 μA, respectively. The CE will transfer the charge carrier from Py to p-Si, which will increase the resistance of Py and reduce the resistance of p-Si. Previously reported charge transfer study[9] showed a MIT like behavior at ~206 K and the resistivity of the p-Si layer was estimated to be ~$1.45\times10^{-5}$ Ωm at 300 K. We calculated that a ~51 nm layer of p-Si with ~$1.45\times10^{-5}$ Ωm average resistivity along with 1% increase in Py resistance will give rise to the observed reduction in the sample resistance from 164.38 Ω to 161.37 Ω while the resistivity of the rest of the p-Si layer remains unchanged. We propose that ~51 nm is the penetration depth of charge carriers due to flexoelectricity mediated CE in 2 μm degenerately doped p-Si. The calculation presented here assumed the resistivity of the charged p-Si based on previously published results. However, any change in the resistivity can lead to a different penetration depth but the order of penetration depth is not expected to be different for similar strain gradient.

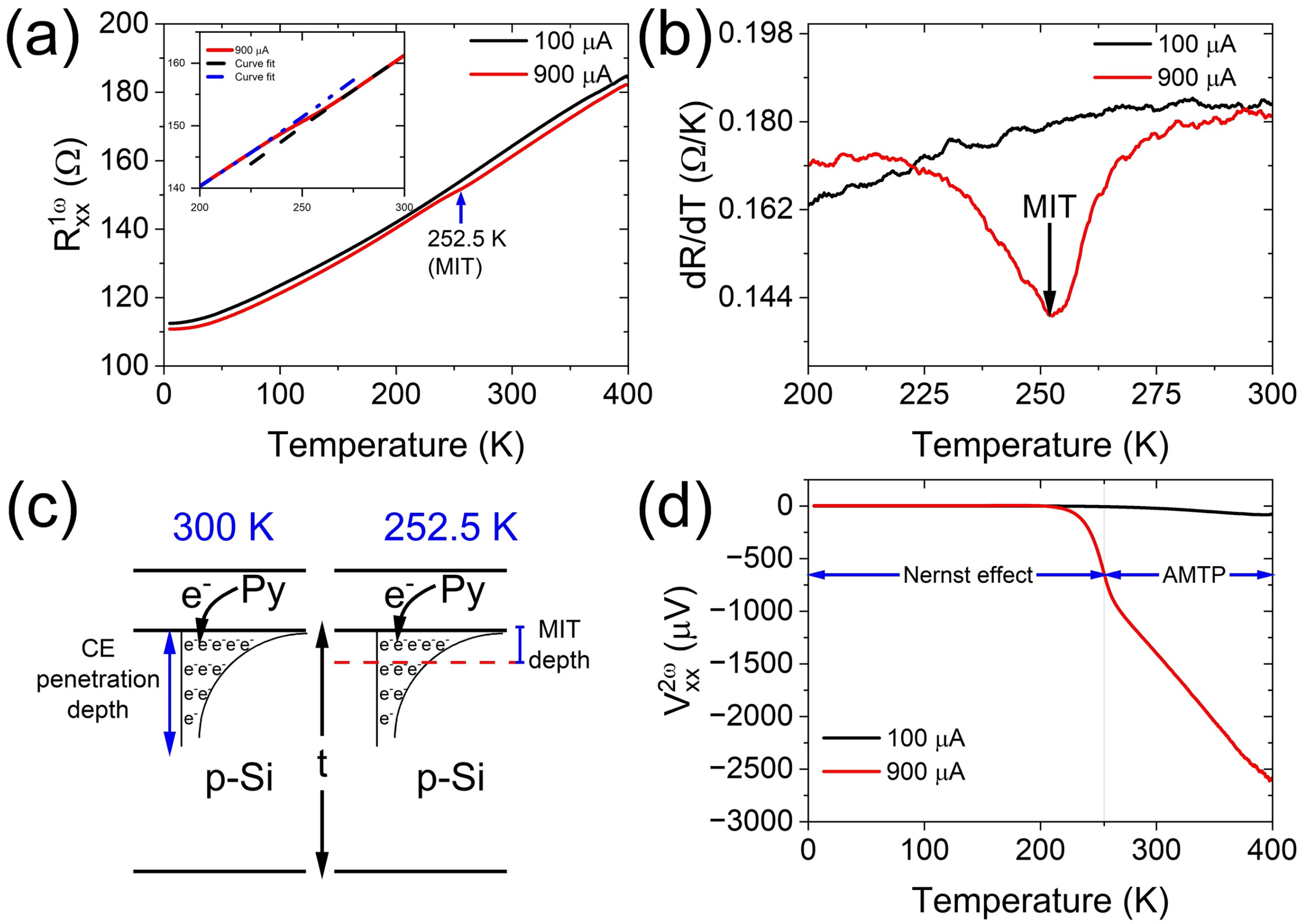

Figure 4. (a) The resistance measurement as a function of temperature from 400 K to 5 K for an applied current bias of 100 μA and 900 μA, while inset shows the polynomial fit in the resistance response at 900 μA above and below the 252.5 K. (b) The $\left(\frac{dR}{dT}\right)$ as a function of temperature between 200 K and 300 K for 100 μA and 900 μA measurements showing MIT at ~252.5 K. (c) A schematic showing the penetration depth and MIT depth due to flexoelectricity mediated CE in Si layer with thickness t, and (d) the $V_{xx}^{2\omega}$ response as a function of temperature measured at 100 μA and 900 μA current bias showing transition from anisotropic magneto-thermopower and Nernst effect behavior.

Then, we analyzed the resistance behavior around the MIT temperature of ~252.5 K. Based on the data fitting above and below the transition as shown in Figure 4 (a) inset, we estimate that resistance must increase from 151 Ω to 152.5 Ω at 252.5 K. However, the resistance measured using 100 μA is 153.65 Ω. If the ~51 nm of charged p-Si layer from CE had a MIT across its whole thickness then resistance after MIT should have been closer to 153.65 Ω and not 152.5 Ω. Hence, only a part of the penetration depth exhibits MIT behavior. We estimate that ~21 nm of the charged p-Si layer near the interface has undergone MIT at 252.5 K. The observed behavior can be explained schematically as shown in Figure 4 (c). At 300 K, the flexoelectricity mediated CE gave rise to interlayer charge transfer where 1% of the charge carriers from Py layer are injected in the p-Si layer. This charge carrier injection achieves a penetration depth of ~51 nm due to screening from the rest of the p-Si, which remains unaffected. The injected charge carrier population is expected to show maximum density near the interface and gradually reaching the doping density around 51 nm away from the interface. Out of ~51 nm, a ~21 nm layer adjacent to the interface, where the charge carrier density of the injected electrons is highest, transitions to an insulator at 252.5 K as observed in the resistance measurement.

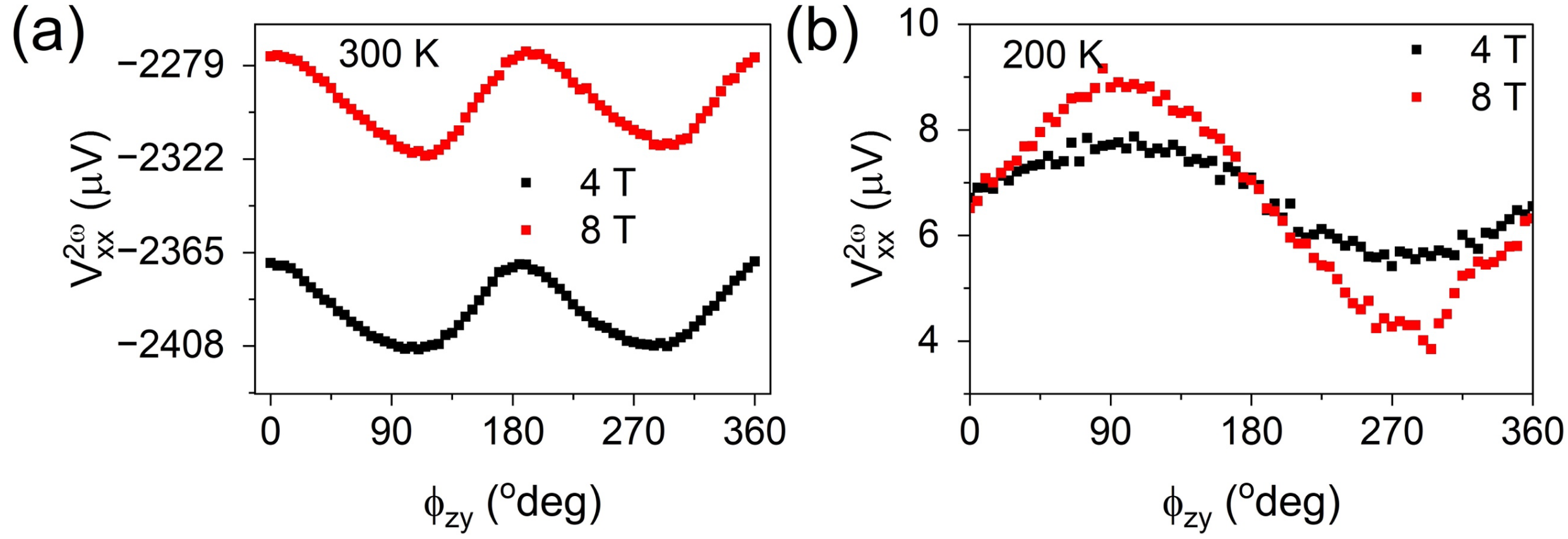

Figure 5. The angle dependent $V_{2\omega}$ response measured as a function of sample angular position in the zy-plane at constant magnetic field (4 T and 8 T) and at (a) 300 K and (b) 200 K.

We have assumed that the change in resistance at 252.5 K is a MIT due to increase in resistance. However, a secondary proof to justify this assumption is required. To address it, we measured the longitudinal second harmonic response as a function of temperature at 100 μA and 900 μA of current bias as shown in Figure 4 (d). In addition, we measured the angle dependent (zy- or cross-sectional plane) longitudinal second harmonic response for a constant magnetic field of 4 T and 8 T at 300 K and 200 K (above and below the MIT) as shown in Figure 5. The longitudinal second harmonic response was very small at 100 μA but significantly larger at 900 μA. For 900 μA measurement, the second harmonic response decrease as a function temperature but shows a change of slope just above the MIT temperature. The angle dependent second harmonic response at 300 K shows a $\sin^2\phi_{zy}$ response, which may arise from anisotropic magneto-thermopower (AMTP). The AMTP response arises due to ferromagnetic proximity effect because the symmetry is same as AMTP response in Py thin films[23], which shows the interlayer coupling due to CE. The AMTP response disappears at 200 K and a $\sin\phi_{zy}$ response is observed, which is attributed to Nernst effect from a temperature gradient along the thickness direction ($\nabla T_z$). This behavior arises due to decoupling of the Py and p-Si layers from 21 nm of MIT layer. A reduction in the charge transfer due to CE can modify the penetration depth as well as MIT response. This change can be achieved using shorter sample length, which will reduce the strain gradient and, as a consequence, the flexoelectricity mediated CE. We did a control experiment, where length of the freestanding sample was 40 μm as compared to 160 μm in this study. We found that AMTP response can be observed even at 20 K when the Py and p-Si layers are coupled in the absence of interfacial MIT layer as shown in Supplementary Figure S1. Hence, the change in resistance at 252.5 K is due to MIT in the charged p-Si layer near the interface. The MIT also led to transition from AMTP behavior to Nernst effect response in the second harmonic response measurement. The experimental results in sample 3 clearly demonstrate CE mediated charge carrier transfer and penetration depth of 51 nm in 2 μm p-Si as hypothesized and schematically shown in Figure 1 (a).

The experiments in sample 1 and 2 demonstrated that flexoelectricity mediated CE leads to charge carrier diffusion across the 400 nm thickness of p-Si layer in Py/p-Si thin film heterostructures. This is attributed to the lack of screening effect from the mobile charge carrier in p-Si. The experimental results on sample 3 demonstrated that CE leads to charge diffusion to a length of ~51 nm in 2 μm thick p-Si layer in Py/p-Si thin film heterostructure. In a 2 μm thick p-Si layer, the screening effect from the mobile charge carrier is relatively strong, which restricts the penetration of charge carrier inside the p-Si layer. A thicker p-Si is expected to restrict the charge carrier accumulation to the interface only. This is a direct experimental proof of length scale effect in flexoelectricity mediated CE as hypothesized and as shown in Figure 1 (a).

In conclusion, we have demonstrated length scale effect in flexoelectricity mediated CE in Py and p-Si heterostructures. The flexoelectricity mediated CE led to charge transfer from Py layer to p-Si layer. When the thickness of the p-Si layer is 2 μm, the charge carrier transfer diffuse from interface to a depth of 51 nm inside the p-Si layer. Whereas the charge carrier transfer diffuse across the 400 nm thickness of a 400 nm thick p-Si layer. The reduction in the electrostatic screening effect leads to the observed behavior, which makes the 400 nm thick p-Si layer electrostatically transparent. The novel results presented in this work may lay the foundation of new CE based methods to control the physical properties in conducting thin film materials similar to electrostatic gating[24].

**Acknowledgement**

The fabrication of experimental devices was done at the Center for Nanoscale Science and Engineering at UC Riverside. The electron microscopy sample preparation and imaging were done at the Central Facility for Advanced Microscopy and Microanalysis at UC Riverside. SK acknowledges research gift from Dr. S Kumar.

**Conflict of interest statement**

The authors declare no competing interests.

**Supplementary Materials-Evidence of length scale effect in contact electrification in conducting thin film heterostructures**

Paul C. Lou[1], Ravindra G. Bhardwaj[2], Anand Katailiha[1], W.P. Beyermann[3] and Sandeep Kumar[4*]

[1] Department of Mechanical Engineering, University of California, Riverside, CA, USA

[2] Department of Mechanical Engineering, Birla Institute of Technology and Science Pilani, Dubai Campus, Dubai, the United Arab Emirates

[3] Department of Physics and Astronomy, University of California, Riverside, CA, USA

[4] Independent researcher, Kurukshetra, India

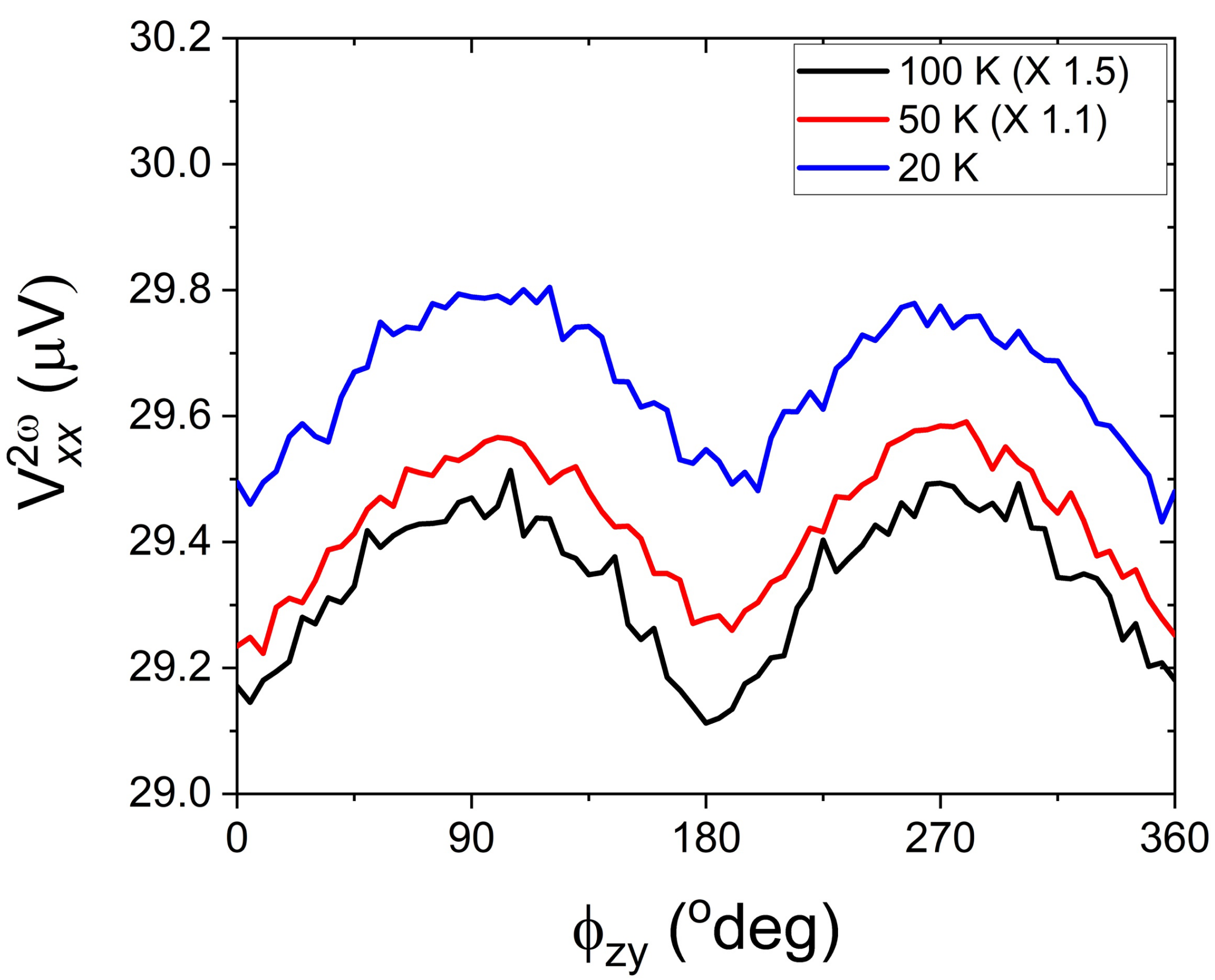

Supplementary Figure S1. The angle dependent $V_{xx}^{2\omega}$ response measured in a Py (25 nm)/MgO (1.8 nm)/p-Si (2 μm) sample (sample length 40 μm) as a function of sample angular position in the zy-plane at constant magnetic field (4 T) and at 100 K, 50 K and 20 K. Since the sample length is smaller, the strain gradient and flexoelectricity mediated contact electrification will also be smaller. As a consequence, no metal-insulator transition was observed and AMTP response is observed at low temperatures. This supports our argument (presented in the main text) of metal-insulator transition in configuration 2 at 252.5 K as shown in Figure 4 and 5.